\documentclass[aps,  prl, twocolumn,showpacs,superscriptaddress, preprintnumbers, nofootinbib]{revtex4-1}  
\usepackage{graphicx}  
\usepackage{dcolumn}   
\usepackage{amsmath}        
\usepackage{amssymb}   
\usepackage{enumerate}
\usepackage{ulem,color}

\newcommand{\tev}{~\text{TeV}}
\newcommand{\gev}{~\text{GeV}}

\begin{document}

\preprint{\textbf{TIFR/TH/15-01}}

\title{New Supersoft Supersymmetry Breaking Operators and a Solution to the $\mu$ Problem}
\author{Ann E.~Nelson}
\affiliation{Department of Physics, University of Washington, Seattle WA, 98195, USA}
\author{Tuhin S.~Roy}
\affiliation{Department of Theoretical Physics, Tata Institute of Fundamental Research, Mumbai 400005, India}
\affiliation{Theory Division T-2, Los Alamos National laboratory, Los Alamos, NM 87545, USA}

\date{\today}

\begin{abstract}
We propose the framework, ``generalized supersoft supersymmetry breaking".  ``Supersoft'' models, with $D$-type supersymmetry breaking and heavy Dirac gauginos,  are considerably  less constrained by the LHC searches than the well studied MSSM. These models also ameliorate the supersymmetric flavor and $CP$ problems. However, previously considered mechanisms for obtaining a natural  size Higgsino mass parameter (namely, $\mu$)   in supersoft models have been relatively complicated and contrived. 
Obtaining  a $125\gev$ for the mass of the lightest Higgs boson has also been difficult. Additional issues with the supersoft scenario arise from the fact that these models contain new scalars in the  adjoint representation of the standard model, which may obtain negative squared-masses, breaking color and generating too large a $T$-parameter.
In this work we introduce new operators into supersoft models which can potentially solve all these issues.  A novel feature of this framework  is that the new $\mu$-term can give unequal masses to the up and down type Higgs fields,  and the Higgsinos can  be much heavier than the Higgs boson without fine-tuning. However, unequal  Higgs and Higgsino masses   also remove some attractive features of supersoft susy. 
\end{abstract}

\maketitle

Supersymmetry (SUSY) at the electroweak scale offers potential solutions  to the gauge hierarchy  and dark matter problems, along with a route towards a Grand Unified Theory (GUT)\footnote{For a comprehensive review see ~\cite{Martin:1997ns}.}. A crucial ingredient  is the presence of the Higgsinos (the superpartners of the Higgs bosons) with masses at the electroweak scale.  At first glance, this does not appear to be a critical issue, since  a supersymmetric Higgs and Higgsino mass term, namely ``$\mu$'',  is allowed.  In fact, issues regarding Higgsino masses are often trivialized by evoking the argument that due to the  nonrenormalization of the superpotential, any value of $\mu$ is technically natural. However, this response does not address the depth of the problem. The $\mu$-parameter needs to be of the order of the electroweak mass scale, which, in a supersymmetric theory, is not an input parameter in the ultra-violet (UV), but is rather generated in the infrared (IR), after the theory is renormalized down to the IR, and is naturally at the scale of the  superpartner masses~\cite{Dine:1981gu,AlvarezGaume:1981wy,Ibanez:1982fr,Dine:1982zb,AlvarezGaume:1983gj}. These masses, in turn, are functions of the two fundamental mass scales of the theory: $(i)$ the scale of the SUSY breaking vacuum expectation value (vev) in the hidden sector, and  $(ii)$ the mass scale associated with the messenger mechanism which connects the hidden sector and the visible sector fields. In models of dynamical supersymmetry  breaking (DSB), the scale of SUSY breaking  is generated via dimensional transmutation~\cite{Witten:1981nf,Dimopoulos:1981au,Dine:1981za,Affleck:1983rr}. The messenger scale is often the Planck scale~\cite{Cremmer:1982en,Chamseddine:1982jx,Barbieri:1982eh,Polchinski:1982an,Ohta:1982wn, Hall:1983iz,Ibanez:1982ee,AlvarezGaume:1983gj}; or the GUT scale~\cite{Dine:1981gu,Polchinski:1982an,Ibanez:1982ee}; or can  be the scale of DSB~\cite{Nelson:1998gp}.  Inclusion of a bare mass term, which is of the order of the  electroweak scale by pure coincidence makes the theory much less elegant and plausible.

The $\mu$-problem is often discussed in the context of the Minimal Supersymmetric Standard Model (MSSM), which is the  most well-studied incarnation of weak scale SUSY. Note that the MSSM is the weak scale effective theory of an underlying supersymmetric theory,  with SUSY being spontaneously broken by the non-zero vev of the $F$-component of a hidden sector chiral superfield. In this framework, a robust solution to the $\mu$ problem is provided by the Giudice-Masiero mechanism~\cite{Giudice:1988yz}, whereby a manifestly supersymmetric higher dimensional operator   involving the Higgs fields and the SUSY breaking hidden sector  superfield  becomes a $\mu$-term. This mechanism assures that  the $\mu$-term is naturally of the order of the superpartner masses. Note that the SUSY breaking terms of the MSSM are known as ``soft"~\cite{Dimopoulos:1981zb,Sakai:1981gr,Hall:1985dx}, because the resulting theory has only logarithmic UV divergences. Such  logarithmic  divergences however mean that the soft terms are sensitive to short distance flavor  and $CP$ violating physics which could potentially lead to problematic flavor changing neutral currents (FCNC)~\cite{Hall:1985dx,Georgi:1986ku}, and new  phases that could make detectible and potentially excessive contributions to electric dipole 
moments~\cite{Ellis:1982tk,Buchmuller:1982ye,Polchinski:1983zd,delAguila:1983kc,Franco:1983xm,Dugan:1984qf}. More recently, the accumulated null observations have put severe constraints on the MSSM, the most serious of which arises from the lack of observation of excess events with jets $+$ missing energy at the LHC. In weak scale SUSY, events with jets $+$ missing energy are produced mostly due to the production of squarks and gluinos, which subsequently decay to jets and the lightest supersymmetric particles (LSPs). These cross-sections are maximized for degenerate squarks and gluinos, which is a generic feature of the MSSM. Within its framework, squarks receive loop suppressed but log enhanced contribution from the gluino mass as the theory is renormalized down to the IR. Except in the case where the squarks start out to be hierarchically heavier than the gauginos at the UV (such as in split-SUSY~\cite{ArkaniHamed:2004fb,Giudice:2004tc,ArkaniHamed:2004yi}), the gluino mass is always comparable to the squark masses in the MSSM.  Satisfying experimental constraints, therefore, requires the raising of the mass scale of all colored particles.  Also note that, because of the restricted form of the Higgs potential in the MSSM, the top squarks are now required to be very heavy, with mass of order a $\tev$ or more in order to obtain $125\gev$  for the mass of the Higgs boson. Since renormalization of the soft Higgs mass-squared term is proportional to the top squark mass, a heavy top squark gives rise to a finely tuned cancellation in the Higgs mass squared parameter.   Thus, in the   MSSM, with SUSY breaking parameters run down from a high scale, SUSY's promise to explain the origin of the weak scale without fine-tuning, is fading in the light of the LHC Higgs discovery and in the absence of any SUSY discovery\cite{Arvanitaki:2013yja,Evans:2013jna,Baer:2014ica}\footnote{Some viable parameter choices may still be considered natural\cite{Berger:2008cq,Feng:2013pwa,Baer:2014ica}, either because of  cancellations in the renormalization group running, or because running from high scales is not considered.}. 

An alternative way to break supersymmetry  is via a vev for the $D$-component of a  hidden sector real superfield~\cite{Fayet:1978qc,Buchmuller:1982ye}. Such  symmetry breaking may be mediated to the  visible sector via a class of operators known as ``supersoft", as they do not induce even logarithmic  ultraviolet divergences in squark and slepton masses~\cite{Fox:2002bu}. The most important previously considered supersoft operators are those giving rise to Dirac gaugino masses~\cite{Fayet:1978qc,Buchmuller:1982ye,Hall:1990hq,Dine:1992yw}.   In  supersoft models the radiatively generated squark and slepton masses are finite,  flavor symmetric, positive, UV insensitive,  and light compared to the gaugino masses~\cite{Fox:2002bu}. Therefore these models additionally avoid the flavor changing neutral current, naturalness, and $CP$ difficulties of the MSSM.  A heavy gluino suppresses processes such as gluino pair production   and squark-gluino production. Also, the pair production of squarks is reduced as the T-channel diagrams involving gluinos do not contribute. Therefore, Dirac masses    allow  for a reduction in the number of events with jets $+$ missing energy for a given squark mass~\cite{Beenakker:1996ch,Kulesza:2008jb,Beenakker:2009ha,Kulesza:2009kq,Beenakker:2011fu,Heikinheimo:2011fk,Kribs:2012gx,Arvanitaki:2013yja,Kribs:2013oda}. The $\mu$-problem is, however, severe in  supersoft models. The Giudice-Masiero mechanism does not work, since SUSY breaking is not mediated  by the $F$-term of a chiral superfield, but by the $D$-term of a real superfield instead. A solution  was proposed in ref.~\cite{Fox:2002bu}, where the conformal compensator generates masses for Higgsinos. To generate the right Higgsino masses, however, this approach requires a conspiracy among the SUSY breaking scale, the messenger scale, and the Planck scale.  One could reintroduce the gauge singlet chiral superfield with an $F$-term and use the Giudice-Masiero mechanism.  However, such a gauge singlet field may lead to  power law UV sensitivity, and  to additional flavor and $CP$ violating SUSY breaking operators; thus  spoiling the supersoft solution to the SUSY FCNC and $CP$ problems~\cite{Fox:2002bu, Kribs:2007ac,Kribs:2010md}.  It is also conceivable to generate  a  $\mu$-term via a supersymmetric vev of a singlet superfield, again bringing in the possibility of new power law divergences in the singlet potential.  If the singlet carries discrete symmetries, then there could be cosmological problems with the production of domain walls associated with breaking of the discrete symmetries. Another potential problem with supersoft models is that the $D$-term contribution to the Higgs quartic coupling vanishes~\cite{Fox:2002bu}, and accommodating a $125\gev$ Higgs becomes difficult. 

In this letter,  we propose a complete and viable framework of weak scale SUSY, namely   ``Generalized Supersoft Supersymmetry," where all SUSY breaking effects are sourced by the $D$-component of a real field/operator from the hidden sector. We include a new class of  D-term mediated soft (but not necessarily supersoft)  operators that allow for a new solution to the $\mu$-problem, restore the Higgs quartic coupling,  and provide considerable modification to supersoft phenomenology.

The visible sector of our  supersoft model includes the superfields of the MSSM, as well as additional chiral superfields  $ \Sigma_{i} $ in the adjoint representation of the SM gauge groups.  The fermionic components of $\Sigma_i$, (namely, $\psi_i$),  will obtain Dirac masses with the gauginos ($\lambda_i$).
Supersymmetry is broken by a $D$-term  of a   hidden sector real superfield $V'$  
\begin{equation}
	\label{eq:1}
	\mathcal{D} \ \equiv  \  \frac{1}{8}  \left\langle D^2  \bar{D}^2 V' \right\rangle \ > \ 0  \, .
\end{equation}
The messenger sector that connects the visible and hidden sector is assumed to be very heavy and we may integrate it out at the messenger  scale $M_m$, which, in turn, could be as high as the Planck scale. The operators generating the gaugino masses are~\cite{Dine:1992yw}:
\begin{equation}
\label{eq:2}
\begin{split}
&	\int d^2 \theta \   \frac{w_{1,i}}{4}   \  \frac{  \bar{D}^2 D^\alpha V'}{M_m}  \ W_{i,\alpha}  \Sigma_{i}  
		\quad  \longrightarrow \quad
	M_{D_i}  \lambda_i \psi_i \, , \\
&	\qquad \qquad \qquad \text{where } \quad  M_{D_i} =   \frac{ w_{1,i} g_i}{\sqrt{2}}   \frac{\mathcal{D} }{M_m} \, . 
\end{split}	
\end{equation}
In the above, $W_{i,\alpha}$ is the field-strength superfield of $i$-th SM gauge group, with $\alpha$ being the spinor index. $M_m$ is the messenger scale, $w_1$ are dimensionless coupling constants, and $D$ and $\bar{D}$ are superderivatives.

An additional class of supersoft terms gives mass to the scalar components of the $\Sigma_i$ fields:
\begin{equation}
\label{eq:3}
\begin{split}
	\int d^2 \theta \   \frac{w_{3,i}}{4}   \  \frac{\left(\frac{1}{4}\bar{D}^2 D V' \right)^2}{M_m^2 }  \Sigma_i^2		\  \longrightarrow \   \left(\frac{ w_{3,i} }{2}   \frac{\mathcal{D}^2 }{M_m^2}  \right) \frac{\sigma_i^2}{2} \, .
\end{split}		
\end{equation}
In Eq.~\eqref{eq:3},  $\sigma_i$ denotes the scalar components of the $\Sigma_i$ chiral superfields.  Since 
these operators are generated at the messenger scale, the scalar masses are of the order of the gaugino masses. Note that even though the gaugino mass operators in Eq.~\eqref{eq:2} give rise to masses for the real components of $\sigma$ fields,  Eq.~\eqref{eq:3} remains the only source of masses for the imaginary components at tree level.
Also, given the fact that the squared-masses generated in   Eq.~\eqref{eq:3} are  linear in the coupling constants $w_{3,i}$, these can be negative, giving rise to nonzero vev for the color octet field, thus breaking color. The gaugino mediated squared-masses for these fields are positive. However, as explained before, these masses are loop suppressed and \emph{not} log-enhanced and are, therefore, small with respect to (w.r.t.) the masses in Eq.~\eqref{eq:3}. In gauge mediated supersoft models, some intricate model building is required to avoid  negative masses squared for some of the adjoint scalars \cite{Fox:2002bu,Benakli:2009mk,Benakli:2010gi,Csaki:2013fla}. Both sets of terms are invariant under the hidden sector gauge symmetry $V'\rightarrow V'+\Lambda+\Lambda^\dagger$, where $\Lambda$ is a chiral superfield.   As discussed in ref.~\cite{Fox:2002bu} this hidden sector gauge invariance is key to the absence of UV sensitive contributions to supersymmetry breaking scalar masses.

In this work we 
propose  a new class of operators which ameliorates all of the previously 
mentioned problems in this framework: 
\begin{equation}
\begin{split}
	- \int d^2 \theta \ \frac{1}{4}  \ w_{2, \Phi_1\Phi_2} \ \frac{\bar{D}^2  \left( D^\alpha V'  D_\alpha \Phi_1 \right)} 	
			{M_m} \Phi_2 
\end{split} \label{eq:4}
\end{equation}
In Eq.~\eqref{eq:4}, $\Phi_{1}$ and $\Phi_{2}$ are visible sector chiral superfields such that the bilinear $\Phi_{1}\Phi_{2}$ is a gauge singlet. Examples of such bilinear gauge singlet in the weak scale supersymmetry are $H_uH_d$, and $\Sigma_i^2$. Note that the operators as expressed in  Eq.~\eqref{eq:4} are manifestly chiral (and part of the superpotential) because of the fact that $\bar{D}^3 = 0$. The terms in Eq.~\eqref{eq:4}  can be given a gauge invariant form (but not supersymmetric), since if $V'$ is set equal to its vev, we find: 
\begin{align}
 \bar{D}^2 \left( D^\alpha  V'   D_\alpha \Phi_1 \right)  =\left(\bar{D}^2 D^\alpha  V'  \right) D_\alpha \Phi_1 + \dots \, , 
 \end{align}
 where $\dots$ represent extra terms that do not contribute to the superpotential.
 When we treat our operators containing $V'$ as a  spurion, since it can come either from a supersymmetric or a gauge invariant operator, it will only generate gauge invariant corrections to SUSY breaking operators, and hence cannot generate terms which require non gauge invariant counter-terms.   There are however other spurionic terms which share the feature of being either supersymmetric or gauge invariant,  which can contribute to squark and slepton masses and nonsupersymmetric trilinears, so the new operators are not necessarily supersoft.    
One important aspect of this operator is that ordering of $\Phi_1$ and $\Phi_2$ in Eq.~\eqref{eq:4} matters in case these represent different fields. 
Expanding Eq.~\eqref{eq:4},  we find masses for all the fermionic components of $\Phi_1$ and $\Phi_2$, and for the  scalar components of $\Phi_2$ only. The scalar components of $\Phi_1$ remain massless.
\begin{equation}
\label{eq:5}
\begin{split}
	& \frac{ \mu_{\phi_2}}{2} \left( \tilde{\phi}_1 \tilde{\phi}_2  - 
		2  F_{\phi_1} \phi_2  \right) \ \rightarrow \ 
			\frac{ \mu_{\phi_2}}{2}   \tilde{\phi}_1 \tilde{\phi}_2 	+ 
				 \left| \mu_{\phi_2} 	\right|^2  \left|\phi_2  \right|^2 \\
	& \quad \text{ where } \quad  \mu_{\phi_2} \ = \  2 \: 
		 w_{2, \Phi_1\Phi_2}  \:  \frac{ \mathcal{ D} }{M_m} \, ,
\end{split} 		
\end{equation}
where $\phi_i$, $\tilde{\phi_i}$, and $F_{\phi_i}$ are the scalar, fermion, and auxiliary components of the chiral multiplet $\Phi_i$ respectively. 

A non-zero value of either  or both of $w_{2, H_u H_d}$, or $w_{2, H_d H_u} $ generates masses for the Higgsinos.
A nice feature of these Higgsino masses is that the masses are naturally of the order of the gaugino masses and are sourced by a single mass scale (\textit{i.e.} vev of the $D$-component of the hidden sector field). These new operators are also phenomenologically important.  Eq.~\eqref{eq:5} implies that unlike the conventional $\mu$ term, $w_{2, H_u H_d} $ only gives rise to down-type Higgs soft masses. The general contributions to the Higgs sector from these unconventional operators (with both  $w_{3, H_u H_d} $ and  $w_{3, H_d H_u} $) are then characterized by not one $\mu$ parameter, but rather by two separate mass parameters (namely, $\mu_u$ and $\mu_d$): 
\begin{equation}
\label{eq:6}
	\frac{1}{2} \left( \mu_u  + \mu_d \right) \tilde H_u  \tilde H_d + 
		 \left|\mu_u \right|^2   \left|h_u\right|^2 + 
    	 		\left|\mu_d \right|^2   \left|h_d\right|^2  \, . 
\end{equation}
Only in the limit $\mu_u = \mu_d = \mu$, the mass terms become identical to that of the conventional $\mu$-term.  A large mass term for $H_d$, will result in large $\tan\beta$ but a potentially natural spectrum. It is, therefore, possible to consider a model in which the Higgsinos and additional scalar bosons are substantially heavier than the Higgs without fine-tuning. This setup also challenges the conventional wisdom regarding fine-tuning in models of weak scale SUSY. Since there is no observable that directly gives a measure of the messenger scale of the theory (and the size of the large logarithmic contribution to the Higgs mass), measuring masses of the Higgsinos seems to be the best way of estimating the size of cancellation needed in order to produce the electroweak scale. Even though exceptions were constructed, where  the cancellation is the result of dynamics~\cite{Roy:2007nz, Murayama:2007ge, Perez:2008ng}, not fine-tuning, (therefore,  the naive interpretation of Higgsino masses being the measure of fine-tuning is incorrect) the belief remains widespread. Eq.~\eqref{eq:6} provides an explicit example, where the Higgsino mass can be made large (because of large $ \mu_d $), without contributing to soft mass of the up-type Higgs.  However, too large  a $\left( \left|\mu_d \right|^2 - \left|\mu_u \right|^2 \right)$, generates  a $\log$-divergent, though loop suppressed Hypercharge $D$-term, which, if too large, can give some scalars tachyonic masses\footnote{We thank Andrew G. Cohen, and Martin Schmaltz for pointing this out to us.}. Also, $\mu_u \neq \mu_d$, can give rise to additional log divergent contributions to scalar soft $\text{masses}^{2}$. For consistency, we assume that all terms which are needed for renormalization are present, and so in the case $\mu_u \neq \mu_d$ squark and slepton masses squared must also receive non supersoft contributions, however such terms can naturally be smaller than the supersoft contributions.

The operator in  Eq.~\eqref{eq:4}, with $\Phi$ replaced by the $\Sigma_i$ fields, can also provide potential solutions associated with the scalar adjoints.  Operators with $w_{3,\Sigma_i^2}$ generate positive definite squared-masses for the scalar components, and  Majorana masses for the fermionic components of the $\Sigma_i$ fields.    
\begin{equation}
\label{eq:7}
\begin{split}
 	\frac{1}{2}M_{N_i}  \psi_i^2 \ + \  \frac{1}{2} \left| M_{N_i} \right|^2  \left| \sigma_i \right|^2 \, ,   \quad
		M_{N_i} = 2 \: w_{2, \Sigma_i^2}  \frac{\mathcal{D} }{M_m}
\end{split}
\end{equation}
Color breaking can be easily avoided (at tree level) for large enough $w_{3,3}$. As mentioned earlier, the gaugino mediated contributions to scalar soft masses at one loop are already positive definite. 

An additional effect of the large masses for the $\sigma$ fields is the (partial) recovery of the Higgs quartic coupling. Take for example, the on-shell Lagrangian \emph{in} the presence of the $\sigma_2$ fields, and the effective Lagrangian after the real components of $\sigma_2$ are integrated out:
\begin{align}
\begin{split}
	\mathcal{L}_{\text{on-shell}} \   \supset \  & \frac{1}{2} \left(  2 M_{D_2} \sigma_{2_R} + \frac{g_2^2}{2} 
			\sum_k  q_k^* t_a q_k \right)^2  \\ 
				 & + \ \quad \frac{1}{2}
		M_{N_2}  \left(   \sigma_{2_R}^2 + \sigma_{2_I}^2  \right) 
\end{split}		\label{eq:9}	\\ 
	\mathcal{L}_{\text{eff}} \   \supset \  & \frac{M_{N_2}^2}{M_{N_2}^2 + 4 M_{D_2}^2} \ 
		\frac{g_2^2}{8}  \sum_k  \bigg( q_k^* t_a q_k \bigg)^2	\, . 
\label{eq:10}
\end{align}
We use the notation $\sigma_{2_R}$ and $\sigma_{2_I}$  to designate the real and  the imaginary parts of $\sigma_2$. Eqs.~(\ref{eq:9}-\ref{eq:10}) are also useful for demonstrating the fact that unlike in the MSSM, $D$-terms of the gauge fields do not contribute to the Higgs quartic in supersoft SUSY. Since the mass term $M_{N_2}$ gets generated only by the operator in Eq.~\eqref{eq:7}, the supersoft limit can be achieved by taking  $M_{N_2} \rightarrow 0$, when  the $D$-term containing the  Higgs quartic vanishes.  In the opposite limit, namely  $M_{N_2} \gg M_{D_2}$, one recovers the full MSSM strength quartic at the tree level. 

The gauginos are no longer Dirac particles once the operators of Eq.~\eqref{eq:7} are included. For instance, the gluinos $\tilde{g}$ and their Dirac partners $\psi_3$ obtain masses from two independent sources:
\begin{align}
		 \mathcal{L}_{\text{gluinos}} \ \supset \  \frac{1}{2} \:
	 	\begin{pmatrix} \tilde{g} & \psi_3 \end{pmatrix} &
		\begin{pmatrix} 0 & M_{D_3} \\ M_{D_3} & M_{N_3}  \end{pmatrix} 
		\begin{pmatrix}  \tilde{g} \\ \psi_3 \end{pmatrix}   
\end{align}
Based on the relative strength of the Dirac mass of gluino and the Majorana mass of $\psi_3$, three qualitatively distinct IR spectra emerge:
\begin{enumerate}[(i).]

\item    $M_{N_3} \gg M_{D_3}:$   The gluino mass matrix has the ``seesaw'' texture. The $\psi_3$ field (in fact, the entire $\Sigma_3$ superfield) is integrated out at the scale $M_{N_3}$. The resultant light gluino (light w.r.t $M_{N_3} $) is a Majorana fermion with a mass  inversely proportional to $M_{N_3}$.  The IR effective theory below $M_{N_3} $ is the MSSM, with an added feature of all scalar masses being still supersoft --  in the sense that these masses do not get big $\log$ contribution from UV scales (although they are  sensitive to $\log M_{N_3}$). 

 \item   $M_{N_3} \ll M_{D_3}:$   Gluinos are ``pseudo-Dirac", with two nearly degenerate Majorana color octet fermions, and a small mass splitting.

\item   $M_{N_3} \sim M_{D_3}:$   Gluinos are mixed Majorana-Dirac~\cite{Kribs:2013eua}, with two Majorana color octet fermions and a mass splitting of order their mass.  The squark--quark--(lighter) gluino  coupling deviates from the usual strong coupling constant ($\alpha_s \rightarrow \alpha_s \cos^2 \theta_g$, where $\theta_g$ is the mixing angle in the gluino mass matrix). The associated squark-gluino production cross-section, for example, thus contains an additional factor of  $\cos^2 \theta_g$  which deviates from $1$ at the leading order.

\end{enumerate}

The neutralino and chargino mass matrices are more complicated, and we leave a complete description  for future work~\cite{GGS2}. Here we make a few remarks. In  supersoft SUSY,  the  gauginos, Higgsinos,  and  additional Higgs bosons can naturally be substantially heavier than the squarks and sleptons without fine-tuning. In fact, a 
charged right handed slepton is often predicted to be the lightest supersymmetric particle (LSP) in supersoft models. 
This, however, is problematic since a stable slepton is not cosmologically viable.  In models with a low messenger scale, the gravitino  becomes the LSP, thereby resolving this issue by allowing the slepton to decay into a lepton and gravitino. Depending on the gravitino mass and the reheating scale after inflation, the gravitino may provide a cold or warm dark matter candidate.

In the scenario we provide, a mostly bino-like Majorana fermion could be the LSP.  If its mass  is close to the mass of the right handed charged sleptons, then  it can  become a thermal relic with the right density due to co-annihilation~\cite{Griest:1990kh}.  Consider the case, where $M_{D_1} \ll M_{N_1}, M_{D_2}, M_{N_2}, \mu_u, \mu_d$.  Since $M_{D_1} \ll M_{N_1}$, there is a potentially  light  mass eigenstate which  is mostly a bino-like Majorana fermion,  which  can be chosen to yield the right thermal relic abundance.
The right-handed charged slepton receives loop suppressed and finite mass which, at one loop,  is of the order of $\left( g_1/2 \pi \right) M_{N_1}^2\log\left(M_{D_1}/ M_{N_1}\right)/M_{D_1} $. We may, without affecting naturalness, add flavor universal soft slepton mass squared terms which are large enough that the right handed slepton mass   is similar in size to the Bino mass.

 In summary, we have shown that adding a new class of  operators   to models with supersoft supersymmetry breaking can offer a solution to the   $\mu$-problem and have very attractive consequences.  
Gluinos in these models can be naturally heavy, several times the mass of the squarks, while the  remainder of the sub TeV  superpartner spectrum can be MSSM like, including the  possibility of WIMP dark matter.  With a heavy gluino this scenario is   less constrained by LHC searches and low energy observables than the MSSM, while still allowing a path towards unification and a dynamical solution to the hierarchy problem.
\section*{Acknowledgements}
This work was supported in part by the US Department of Energy under grant
DE-SC0011637. 


\bibliography{supersoft}

\end{document}